\documentclass{cpbtex}
\usepackage{lineno,hyperref,amssymb,amsmath,epsfig,mathrsfs,subfigure}
\begin{document}
\begin{CJK*}{GBK}{song}

\title{On multi-soliton solutions to the Heisenberg ferromagnetic spin chain
equation in (2+1)-dimensions\thanks{This work is supported by the National Natural Science Foundation of China (Grant No.~61775126).}}


\author{Zhou-Zheng Kang\thanks{Corresponding author. E-mail:~zhzhkang@126.com}, \ Rong-Cao Yang\\
{\small College of Physics and Electronic Engineering, Shanxi University, Taiyuan 030006, China}\\  
}   


\date{}
\maketitle

\begin{abstract}
This paper concentrates on the Heisenberg ferromagnetic spin chain (HFSC) equation in (2+1)-dimensions modelling nonlinear wave propagation in ferromagnetic
spin chain. A variable transformation is first employed to reduce the studied equation. And then an associated matrix Riemann-Hilbert problem is built on the real line through analyzing spectral problem of the reduced equation. As a consequence,
solving the obtained matrix Riemann-Hilbert problem with the identity jump matrix, corresponding to the reflectionless, the general multi-soliton solutions to the HFSC equation in (2+1)-dimensions are acquired. Specially, the one- and two-soliton solutions are worked out and analyzed graphically.

\end{abstract}

\textbf{Keywords:} Heisenberg ferromagnetic spin chain equation; Riemann-Hilbert problem; soliton solutions


\section{Introduction}
Wave phenomena appear in many fields of science and engineering, such as fluid mechanics, optics, solid mechanics, electromagnetism, structural mechanics and quantum mechanics. The waves for these applications are described by solutions to nonlinear partial differential equations (NLPDEs).
To date, a number of methods have confirmed the effectiveness in investigating solutions to NLPDEs, some of which include the inverse scattering transformation$^{[1]}$, the Hirota bilinear method$^{[2]}$, the Darboux transformation$^{[3]}$ and the B\"acklund
transformation$^{[4]}$. In recent years, researchers have shown an increased interest in study of NLPDEs utilizing the Riemann-Hilbert technique, such as the coupled nonlinear Schr\"odinger equation$^{[5]}$, the Kundu--Eckhaus equation$^{[6]}$, the coupled mKdV system$^{[7]}$ and the $N$-coupled Hirota equation$^{[8]}$.

In this letter, we consider the Heisenberg ferromagnetic spin chain equation$^{[9]}$ in (2+1)-dimensions
\begin{equation}
iu_{t}+\alpha_{{1}}u_{{{\it xx}}}+\alpha_{{2}}u_{
{{\it yy}}}+\alpha_{{3}}u_{{{\it xy}}}-\alpha_{{4}}{|u|}^{2}u=0,
\end{equation}
where $\alpha_{{1}}={\kappa}^{4}(\mu+\mu_{2}),\alpha_{{2}}={\kappa}^{4}(\mu_{1}+\mu_{2}),\alpha_{{3}}=2{\kappa}^{4}\mu_{2}$ and $\alpha_{{4}}=2{\kappa}^{4}\mu_{3}.$
In Eq. (1), $\kappa$ is a lattice parameter, $\mu$ and $\mu_{1}$ are the coefficients of bilinear exchange interactions along $X$- and $Y$-directions, $\mu_{2}$ means a neighboring interaction factor along the diagonal, and $\mu_{3}$ is an uniaxial crystal field anisotropy parameter.
Equation (1) is integrable and model nonlinear wave propagation in ferromagnetic spin chain$^{[10]}$.
To date, many studies have been conducted on Eq. (1) and lots of results$^{[10-24]}$ have been achieved. For instance, Lax pair was presented, based on which and Darboux transformation, a series of rogue wave solutions$^{[16]}$ were attained. In a follow-up study,
the mixed breather and rogue wave solution was explored by extended Darboux technique, and the interaction behaviors between the mixed waves were analyzed in detail$^{[20]}$.
The main goal of the current study is to determine multi-soliton solutions to Eq. (1) in the framework of Riemann-Hilbert problem.

\section{Matrix Riemann-Hilbert problem}
Our starting point is the transformation $\tilde{x}=x+ky$, which can reduce Eq. (1) to the following equation
\begin{equation}
iu_{t}+(\alpha_{{1}}+\alpha_{{2}}k^{2}+\alpha_{{3}}k)u_{{{\it \tilde{x}\tilde{x}}}}-\alpha_{{4}}{|u|}^{2}u=0.
\end{equation}
The Lax pair$^{[16]}$ for Eq. (2) reads
\begin{align}
&{{\varphi}_{\tilde{x}}}=U\varphi=(i\varsigma \Lambda+iQ)\varphi,\\
&{{\varphi}_{t}}=V\varphi=(-i\alpha_{4}{\varsigma}^{2}\Lambda+V_{1})\varphi,
\end{align}
where ${\varphi}={({\varphi}_{1},{\varphi}_{2})^\textrm{T}}$ is the spectral function, the superscript T stands for the vector transpose, and $\varsigma\in \mathbb{C}$ is a spectral parameter. Moreover, $\alpha_{4}=-2(\alpha_{{1}}+\alpha_{{2}}k^{2}+\alpha_{{3}}k),\Lambda =\text{diag}(1,-1)$,
$$
Q=\left(
    \begin{matrix}
      0 & u^{*} \\
      u & 0 \\
    \end{matrix}
  \right),\quad
V_{1}=\left(
    \begin{matrix}
      \frac{1}{2}i\alpha_{4}uu^{*} & -i\alpha_{4}\varsigma {u^{*}}-\frac{1}{2}\alpha_{4}u_{\tilde{x}}^{*} \\
      -i\alpha_{4}\varsigma {u}+\frac{1}{2}\alpha_{4}u_{\tilde{x}} & -\frac{1}{2}i\alpha_{4}uu^{*} \\
    \end{matrix}
  \right).$$


In our analysis, the potential $u$ is assumed to rapidly vanish
at very large distances. It is known from (3) and (4) that $\varphi \propto{{{e}}^{i\varsigma\Lambda \tilde{x}-i\alpha_{4}{{\varsigma}^{2}}\Lambda t}}.$
Thus we introduce the transformation
$
\varphi=\psi{{{e}}^{i\varsigma\Lambda \tilde{x}-i\alpha_{4}{{\varsigma}^{2}}\Lambda t}},
$
which helps to turn the Lax pair (3) and (4) into
\begin{align}
&{\psi_{\tilde{x}}}=i\varsigma[\Lambda,\psi]+U_{1}\psi,\\
&{\psi_{t}}=-i\alpha_{4}{{\varsigma}^{2}}[\Lambda,\psi]+V_{1}\psi,
\end{align}
where $[\Lambda,\psi]=\Lambda \psi-\psi\Lambda$ and $U_{1}=iQ$.

Below, we concentrate on the spectral problem (5) to perform spectral analysis. And
$t$ will be viewed as a dummy variable. We express the matrix Jost solutions ${{\psi}_{\pm}}(\tilde{x},\varsigma)$ as
\begin{equation}
{{\psi}_{\pm}}(\tilde{x},\varsigma)=({{[{\psi_{\pm}}]_{1}}},{{[{\psi_{\pm}}]_{2}}})(\tilde{x},\varsigma),
\end{equation}
with the asymptotics
\begin{equation}
{\psi_{\pm}}(\tilde{x},\varsigma)\to \mathbf{I}_{2},\quad \tilde{x}\to \pm\infty ,
\end{equation}
where the subscripts in $\psi$ refer to which end of the $\tilde{x}$-axis the boundary conditions are set, and
$\mathbf{I}_{2}$ is the identity matrix of size 2. By use of the variation of parameters and (8), one can obtain Volterra integral equations
that can be cast in $\psi_{\pm}$ as
\begin{align}
&{\psi_{-}}(\tilde{x},\varsigma)=\mathbf{I}_{2}+\int_{-\infty }^{\tilde{x}}{{{{e}}^{i\varsigma \Lambda (\tilde{x}-z)}}U_{1}(z){\psi_{-}}(z,\varsigma){{{e}}^{-i\varsigma \Lambda (\tilde{x}-z)}}{d}z},\\
&{\psi_{+}}(\tilde{x},\varsigma)=\mathbf{I}_{2}-\int_{\tilde{x}}^{+\infty }{{{{e}}^{i\varsigma \Lambda (\tilde{x}-z)}}U_{1}(z){\psi_{+}}(z,\varsigma){{{e}}^{-i\varsigma \Lambda (\tilde{x}-z)}}{d}z}.
\end{align}
Then, Eqs. (9) and (10) are analyzed to see that ${{[{\psi_{-}}]_{1}}}$ and ${{[{\psi_{+}}]_{2}}}$ allow analytical extensions to ${\mathbb{C}_{-}}$ and continuous for
$\varsigma\in {\mathbb{C}_{-}}\cup \mathbb{R}$, however, ${{[{\psi_{+}}]_{1}}}$ and ${{[{\psi_{-}}]_{2}}}$ are analytically extendible to ${\mathbb{C}_{+}}$ and continuous for $\varsigma\in {\mathbb{C}_{+}}\cup \mathbb{R}$, where ${\mathbb{C}_{-}}$
and ${\mathbb{C}_{+}}$ are the lower and upper half $\varsigma$-plane.

Applying the Abel's identity, it is revealed that $\det{\psi_{\pm}}$ are independent
of $\tilde{x}$, since $\text{tr}Q=0$. Evaluating $\det{\psi_{-}}$ at $\tilde{x}=-\infty$ and $\det{\psi_{+}}$ at $\tilde{x}=+\infty$, we know that $\det{\psi_{\pm}(\tilde{x},\varsigma)}=1$ for $\forall \tilde{x}$ and
$\varsigma \in \mathbb{R}$. Since ${\psi_{-}}{{{e}}^{i\varsigma \Lambda \tilde{x}}}$ and ${\psi_{+}}{{{e}}^{i\varsigma \Lambda \tilde{x}}}$ are matrix solutions of (3), they must be linearly related by the scattering matrix $S(\varsigma)$
\begin{equation}
{{\psi}_{-}}{{{e}}^{i\varsigma \Lambda \tilde{x}}}={\psi_{+}}{{{e}}^{i\varsigma \Lambda \tilde{x}}}S(\varsigma),\quad S(\varsigma)=\left(
                                                                                       \begin{matrix}
                                                                                         s_{11} & s_{12} \\
                                                                                         s_{21} & s_{22} \\
                                                                                       \end{matrix}
                                                                                     \right),
\quad \varsigma \in \mathbb{R}.
\end{equation}
We point out that $\det{S(\varsigma)}=1$ because of $\det{\psi_{\pm}(\tilde{x},\varsigma)}=1$.

A matrix Riemann-Hilbert problem required is related to two matrix analytic functions. In consideration of the analytic properties of $\psi_{\pm}$, we define the analytic function in ${\mathbb{C}_{+}}$ as
\begin{equation}
{{P}_{1}}(\tilde{x},\varsigma)=({{[{\psi_{+}}]_{1}}},{{[{\psi_{-}}]_{2}}})(\tilde{x},\varsigma)=\psi_{+}{H}_{1}+\psi_{-}{H}_{2},
\end{equation}
where
\begin{equation}
{H}_{1}=\text{diag}(1,0),\quad {H}_{2}=\text{diag}(0,1).
\end{equation}

Now, we examine the large-$\varsigma$ asymptotic behavior of ${{P}_{1}}$.
Since ${{P}_{1}}$ solves (5),
we make an asymptotic expansion for ${{P}_{1}}$ at large-$\varsigma$
\begin{equation}
{{P}_{1}}=P_{1}^{(0)}+{\varsigma}^{-1}P_{1}^{(1)}+{{\varsigma}^{-2}}P_{1}^{(2)}+O\big({{\varsigma}^{-3}}\big),\quad \varsigma \to \infty ,
\end{equation}
and plug the asymptotic expansion into (5). Comparing the coefficients of the same powers of $\varsigma$ yields
$$O(1):P_{1\tilde{x}}^{(0)}=i\big[\Lambda,P_{1}^{(1)}\big]+\tilde{U}P_{1}^{(0)};\quad O(\varsigma): i\big[\Lambda,P_{1}^{(0)}\big]=0.$$
Therefore, we find that $P_{1}^{(0)}=\mathbf{I}_{2}$. This means that
${{P}_{1}}\to \mathbf{I}_{2}$ as $\varsigma \in {\mathbb{C}_{+}}\to \infty.$

To determine the analytic counterpart of $P_{2}$ in ${\mathbb{C}_{-}}$, we consider the adjoint equation of (6)
\begin{equation}
{{\chi}_{\tilde{x}}}=i\lambda[\Lambda,\chi]-\chi U_{1}.
\end{equation}
It can be verified that the inverse matrices
\begin{equation}
\psi_{\pm}^{-1}=\left( \begin{matrix}
   {[\psi_{\pm}^{-1}]^{1}}  \\
   {[\psi_{\pm}^{-1}]^{2}}  \\
\end{matrix} \right)
\end{equation}
solve (15),
where $[\psi_{\pm}^{-1}]^{j}(j=1,2)$ denote the $j$-th row of $\psi_{\pm}^{-1}$,
and obey the boundary conditions $\psi_{\pm}^{-1}(\tilde{x},\varsigma)\rightarrow \mathbf{I}_{2}$ as $\tilde{x}\rightarrow\pm\infty.$ It follows from (11) that
\begin{equation}
\psi_{-}^{-1}={{{e}}^{i\varsigma \Lambda \tilde{x}}}S^{-1}(\varsigma){{{e}}^{-i\varsigma \Lambda \tilde{x}}}\psi_{+}^{-1},\quad \varsigma \in \mathbb{R},
\end{equation}
where $S^{-1}(\varsigma)={{({{r}_{lk}})}_{2\times 2}}$. Thus, the analytic function $P_{2}$ in ${\mathbb{C}_{-}}$ is expressed as
\begin{equation}
{{P}_{2}}(\tilde{x},\varsigma)=\left( \begin{matrix}
   {[\psi_{+}^{-1}]^{1}}  \\
   {[\psi_{-}^{-1}]^{2}}  \\
\end{matrix} \right)(\tilde{x},\varsigma)=H_{1}\psi_{+}^{-1}+H_{2}\psi_{-}^{-1},
\end{equation}
with $H_{1}$ and $H_{2}$ being given by (13).
And the large-$\varsigma$ asymptotic behavior of $P_{2}$ is ${{P}_{2}}\to \mathbf{I}_{2}$ as $\varsigma\to \infty .$

Insertion of (7) into (11) yields
$$
{{[{\psi_{-}}]_{2}}}={{s}_{12}}{{{e}}^{2i\varsigma \tilde{x}}}{{[{\psi_{+}}]_{1}}}+{{s}_{22}}{{[{\psi_{+}}]_{2}}}.
$$

Carrying (16) into (17) leads to
$$
{[\psi_{-}^{-1}]^{2}}={{r}_{21}}{{{e}}^{-2i\varsigma \tilde{x}}}{[\psi_{+}^{-1}]^{1}}+{{r}_{22}}{[\psi_{+}^{-1}]^{2}}.
$$

Consequently, $P_{1}$ and $P_{2}$ can be given as
$${{P}_{1}}=({{[{\psi_{+}}]_{1}}},{{[{\psi_{-}}]_{2}}})=({{[{\psi_{+}}]_{1}}},{{[{\psi_{+}}]_{2}}})\left( \begin{matrix}
   1 & {{s}_{12}}{{{e}}^{2i\varsigma \tilde{x}}}\\
   0 & {{s}_{22}}\\
\end{matrix} \right),\quad
{{P}_{2}}=\left(\begin{matrix}
   {[\psi_{+}^{-1}]^{1}}  \\
   {[\psi_{-}^{-1}]^{2}}  \\
\end{matrix}\right)=\left(\begin{matrix}
   1 & 0  \\
   {{r}_{21}}{{{e}}^{-2i\varsigma \tilde{x}}} & {{r}_{22}}\\
\end{matrix}\right)\left(\begin{matrix}
   {[\psi_{+}^{-1}]^{1}}  \\
   {[\psi_{+}^{-1}]^{2}}  \\
\end{matrix}\right).
$$

Up to now, we have presented the analytic functions $P_{1}$ in ${\mathbb{C}_{+}}$ and $P_{2}$ in ${\mathbb{C}_{-}}$, respectively.
Through denoting that ${P_{1}}\rightarrow{P^{+}}$ as $\varsigma\in {\mathbb{C}_{+}}\rightarrow\mathbb{R}$ and ${P_{2}}\rightarrow{P^{-}}$ as $\varsigma\in {\mathbb{C}_{-}}\rightarrow\mathbb{R}$, a matrix Riemann-Hilbert problem required can be stated on the real line as follows
\begin{equation}
{{P}^{-}}(\tilde{x},\varsigma){{P}^{+}}(\tilde{x},\varsigma)=G(\tilde{x},\varsigma)=\left(\begin{matrix}
   1 & {{s}_{12}}{{{e}}^{2i\varsigma \tilde{x}}}  \\
   {{r}_{21}}{{{e}}^{-2i\varsigma \tilde{x}}} & 1  \\
\end{matrix}\right),\quad \varsigma\in \mathbb{R},
\end{equation}
with canonical normalization conditions
$
{{P}_{1}}(\tilde{x},\varsigma)\to \mathbf{I}_{2}$ as $\varsigma \in {\mathbb{C}_{+}}\to \infty$ and
${{P}_{2}}(\tilde{x},\varsigma)\to \mathbf{I}_{2}$ as $\varsigma \in {\mathbb{C}_{-}}\to \infty.
$

In what follows, we present reconstruction formula of the potential. Since $P_{1}(\tilde{x},\varsigma)$ solves (5), expanding
$P_{1}(\tilde{x},\varsigma)$ at large-$\varsigma$ as
$$
{{P}_{1}}(\tilde{x},\varsigma)=\mathbf{I}_{2}+{\varsigma}^{-1}P_{1}^{(1)}+{{\varsigma}^{-2}}P_{1}^{(2)}+O\big({{\varsigma}^{-3}}\big),\quad \varsigma\to \infty,
$$
and plugging this expansion into (5), we see that
$$
Q=-\big[\Lambda ,P_{1}^{(1)}\big]=\left(\begin{matrix}
   0 & -2\big(P_{1}^{(1)}\big)_{12}  \\
   2\big(P_{1}^{(1)}\big)_{21} & 0  \\
\end{matrix} \right)\Longrightarrow u=2{{\big(P_{1}^{(1)}\big)_{21}}}.
$$
Here ${{\big(P_{1}^{(1)}\big)_{21}}}$ stands for the (2,1)-element of $P_{1}^{(1)}$. Hence, the reconstruction for the potential is completed.

\section{Soliton solutions}
Our focus in this section will be on generating soliton solutions to Eq. (1) on basis of the established matrix Riemann-Hilbert problem.
Now suppose that the Riemann-Hilbert problem (19) is irregular, i.e., $\det {{P}_{1}}(\varsigma)$ and $\det {{P}_{2}}(\varsigma)$ can be zeros at certain discrete locations in analytic domains. From $\det{\psi_{\pm}}=1$, (12) and (18), and the scattering relation between ${\psi_{+}}$ and ${\psi_{-}}$, we derive
$
\det {{P}_{1}}(\varsigma)={s_{22}}(\varsigma)$ and $\det {{P}_{2}}(\varsigma)={r_{22}}(\varsigma).
$
Therefore, $\det {{P}_{1}}(\varsigma)$ and $\det {{P}_{2}}(\varsigma)$ are in possession of the same zeros as ${s}_{22}(\varsigma)$ and ${r}_{22}(\varsigma)$. 

With above analysis, we now specify the locations of zeros. Manifestly, the matrix $U_{1}$ possesses the property $U_{1}^{\dagger }=-U_{1}$, (the symbol $\dagger$ here means the matrix Hermitian). Using this property, we deduce
\begin{equation}
\psi_{\pm }^{\dagger }({\tilde{x}},{\varsigma}^{*})=\psi_{\pm }^{-1}(\tilde{x},\varsigma).
\end{equation}
Taking the Hermitian to (12) and using (18), we find that
\begin{equation}
P_{1}^{\dagger }({{\varsigma}^{*}})={{P}_{2}}(\varsigma),\quad \varsigma \in {\mathbb{C}_{-}},
\end{equation}
and the involution property ${{S}^{\dagger }}({{\varsigma}^{*}})={{S}^{-1}}(\varsigma).$ It follows at once that
$
s_{22}^{*}({{\varsigma}^{*}})={{r}_{22}}(\varsigma),
$
which tells us that each zero ${\varsigma_{j}}$ of $\det {{P}_{1}}$ can generate each zero $\varsigma_{j}^{*}$ of $\det{{P}_{2}}$. Let $N\in \mathbb{N}$ be arbitrary. In the generic case,
assume that $\det {{P}_{1}}$ and $\det{{P}_{2}}$ respectively possess simple zeros at ${{\varsigma}_{j}}\in{\mathbb{C}_{+}}$ and ${{\hat{\varsigma}}_{j}}\in{\mathbb{C}_{-}}$, where ${{\hat{\varsigma}}_{j}}=\varsigma_{j}^{*},1\leq j\leq N.$ In this case, each of the kernel of ${{P}_{1}}({{\varsigma}_{j}})$ comprises only a single basis column vector ${{\omega}_{j}}$, and each of the kernel of ${{P}_{2}}({{\hat{\varsigma}}_{j}})$ comprises only a single basis row vector ${{\hat{\omega}}_{j}}$:
\begin{align}
&{{P}_{1}}({{\varsigma}_{j}}){{\omega}_{j}}=0,\\
&{{\hat{\omega}}_{j}}{{P}_{2}}({{\hat{\varsigma}}_{j}})=0,
\end{align}
where ${{\omega}_{j}}$ and ${{\hat{\omega}}_{j}}$ are column and row vectors.
Taking the Hermitian to (22) and utilizing (21), we see that
\begin{equation}
{{\hat{\omega}}_{j}}=\omega_{j}^{\dagger },\quad 1\le j\le N.
\end{equation}
Then taking $\tilde{x}$-derivative and $t$-derivative in (22) respectively, and using (5) and (6), we have
$${{P}_{1}}({{\varsigma}_{j}})\left( \frac{\partial {{\omega}_{j}}}{\partial \tilde{x}}-i{{\varsigma}_{j}}\Lambda {{\omega}_{j}} \right)=0,\quad
{{P}_{1}}({{\varsigma}_{j}})\left( \frac{\partial {{\omega}_{j}}}{\partial t}+i{\alpha_{4}}\varsigma_{j}^{2}\Lambda{{\omega}_{j}} \right)=0.$$
Through computing, we attain
$
{{\omega}_{j}}={{{e}}^{\left(i{{\varsigma}_{j}}\tilde{x}-i\alpha_{4}\varsigma_{j}^{2}t\right)\Lambda }}{{\omega}_{j0}},
$
with ${{\omega}_{j0}}$ being independent of $\tilde{x}$ and $t$. Recalling the relation (24), we obtain
${{\hat{\omega}}_{j}}=\omega_{j0}^{\dagger }{{{e}}^{\left(-i\varsigma _{j}^{*}\tilde{x}+i\alpha_{4}\varsigma {{_{j}^{*}}^{2}}t\right)\Lambda}},1\le j\le N.$

For derivation of soliton solutions, we take $G=\mathbf{I}_{2}$ in (19), corresponding to the reflectionless. Hence, the solutions to the special Riemann-Hilbert problem [25] can be derived as
\begin{equation} {{P}_{1}}(\varsigma)=\mathbf{I}_{2}-\sum\limits_{k=1}^{N}{\sum\limits_{j=1}^{N}{\frac{{{\omega}_{k}}{{{\hat{\omega}}}_{j}}{{\big({{M}^{-1}}\big)_{kj}}}}{\varsigma -{{{\hat{\varsigma}}}_{j}}}}}, \quad {{P}_{2}}(\varsigma)=\mathbf{I}_{2}+\sum\limits_{k=1}^{N}{\sum\limits_{j=1}^{N}{\frac{{{\omega}_{k}}{{{\hat{\omega}}}_{j}}{{\big({{M}^{-1}}\big)_{kj}}}}{\varsigma-{{\varsigma }_{k}}}}},
\end{equation}
with $M$ being an ${N\times N}$ matrix determined by
\begin{equation*}
{{m}_{kj}}=\frac{{\hat{\omega}_{k}}{{{{\omega}}}_{j}}}{{{\varsigma}_{j}}-{{{\hat{\varsigma}}}_{k}}},\quad 1\le k,j\le N.
\end{equation*}

Consequently, by incorporating the establised formulae after involution properties with ${{\omega}_{j0}}={({a_{j}},{b_{j}})^\textrm{T}}$ and
${{\vartheta}_{j}}=i{{\varsigma}_{j}}(x+ky)-i\alpha_{4}\varsigma_{j}^{2}t$, we acquire the explicit expression of general $N$-soliton solution to Eq. (1):
\begin{equation}
u(x,y,t)=-2\sum\limits_{k=1}^{N}{\sum\limits_{j=1}^{N}{b _{k}{{a}_{j}^{*}}{{{e}}^{{\vartheta_{j}^{*}-{\vartheta}_{k}}}}{{\big({{M}^{-1}}\big)_{kj}}}}},\quad
{{m}_{kj}}=\frac{1}{{{\varsigma}_{j}}-\varsigma_{k}^{*}}{\big(a_{k}^{*}{a_{j}}{{{e}}^{\vartheta_{k}^{*}+{{\vartheta}_{j}}}}+b_{k}^{*}{ b_{j}}{{{e}}^{-\vartheta_{k}^{*}-{{\vartheta }_{j}}}}\big)}.
\end{equation}

Our main concern in the rest of this section is to compute the one- and two-soliton solutions.

(i) When $N=1$, a direct computation generates the following one-soliton solution
\begin{equation}
u(x,y,t)=-\frac{2{{a}_{1}^{*}}b_{1}({{\varsigma}_{1}}-\varsigma_{1}^{*}){{{e}}^{\vartheta_{1}^{*}-{{\vartheta}_{1}}}}}{{{\left| {a_{1}} \right|}^{2}}{{{e}}^{\vartheta_{1}^{*}+{{\vartheta}_{1}}}}+{{\left| {b_{1}} \right|}^{2}}{{{e}}^{-\vartheta_{1}^{*}-{{\vartheta}_{1}}}}},
\end{equation}
where ${{\vartheta}_{1}}=i{{\varsigma}_{1}}(x+ky)-i\alpha_{4}\varsigma_{1}^{2}t$.
Upon setting ${b_{1}}=1,{{\left| {a_{1}} \right|}^{2}}={{{e}}^{2{{\xi }_{1}}}}$ and ${{\varsigma}_{1}}={\varsigma_{11}}+i{\varsigma_{12}}$, the solution (27) becomes
\begin{equation}
u(x,y,t)=-2i{{a}_{1}^{*}}{\varsigma_{12}}{{{e}}^{-{{\xi}_{1}}}}{{{e}}^{{{\vartheta}_{1}^{*}}-{\vartheta_{1}}}}\text{sech}(\vartheta_{1}^{*}+{{\vartheta}_{1}}+{{\xi }_{1}}),
\end{equation}
where
$
\vartheta_{1}^{*}+{\vartheta}_{1}=-2\varsigma _{12}(x+ky)+4\alpha_{4}\varsigma _{11}\varsigma _{12}t$ and $
\vartheta_{1}^{*}-{\vartheta}_{1}=-2i\varsigma _{11}(x+ky)+2i\alpha_{4}{\varsigma_{11}^{2}}t-2i\alpha_{4}{\varsigma _{12}^{2}}t.
$
Equivalently, the solution (28) reads
\begin{equation}
u(x,y,t)=-2i{{a}_{1}^{*}}{\varsigma_{12}}{{{e}}^{-{{\xi }_{1}}}}{{{e}}^{-2i\varsigma _{11}(x+ky)+2i\alpha_{4}{\varsigma_{11}^{2}}t-2i\alpha_{4}{\varsigma _{12}^{2}}t}}\text{sech}(-2\varsigma _{12}(x+ky)+4\alpha_{4}\varsigma _{11}\varsigma _{12}t+{{\xi }_{1}}).
\end{equation}

It is pointed out from (29) that the amplitude function $|u|$ is a sech-shaped solitary wave with peak amplitude $2|{{a}_{1}^{*}}|{\varsigma_{12}}{{{e}}^{-{{\xi }_{1}}}}.$ The phase linearly relys on the spatial variables $x,y$ and temporal variable $t$.
This wave propagates at velocity $2\alpha_{4}\varsigma _{11}$ after setting $y=0$, which is merely dependant on the real part of the spectral parameter ${\varsigma}_{1}$.
By choosing parameters as $a_{1}=1,b_{1}= 0.5,\varsigma _{11}= 0.2,\varsigma _{12}=0.3,\xi_{1}=0,k=1,\alpha_{1}=1,\alpha_{2}=1,\alpha_{3}=1,y=0$, then the localization of this solution is plotted in $(x,t)$-plane. From Fig. 1(b), it is seen that the wave travels towards the negative direction of the $x$-axis as time evolves.
\begin{figure}
\begin{center}
\subfigure[]{\resizebox{0.34\hsize}{!}{\includegraphics*{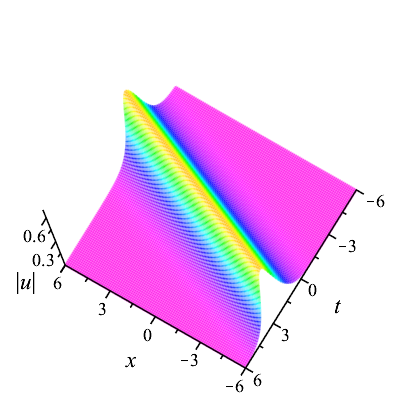}}}
\subfigure[]{\resizebox{0.29\hsize}{!}{\includegraphics*{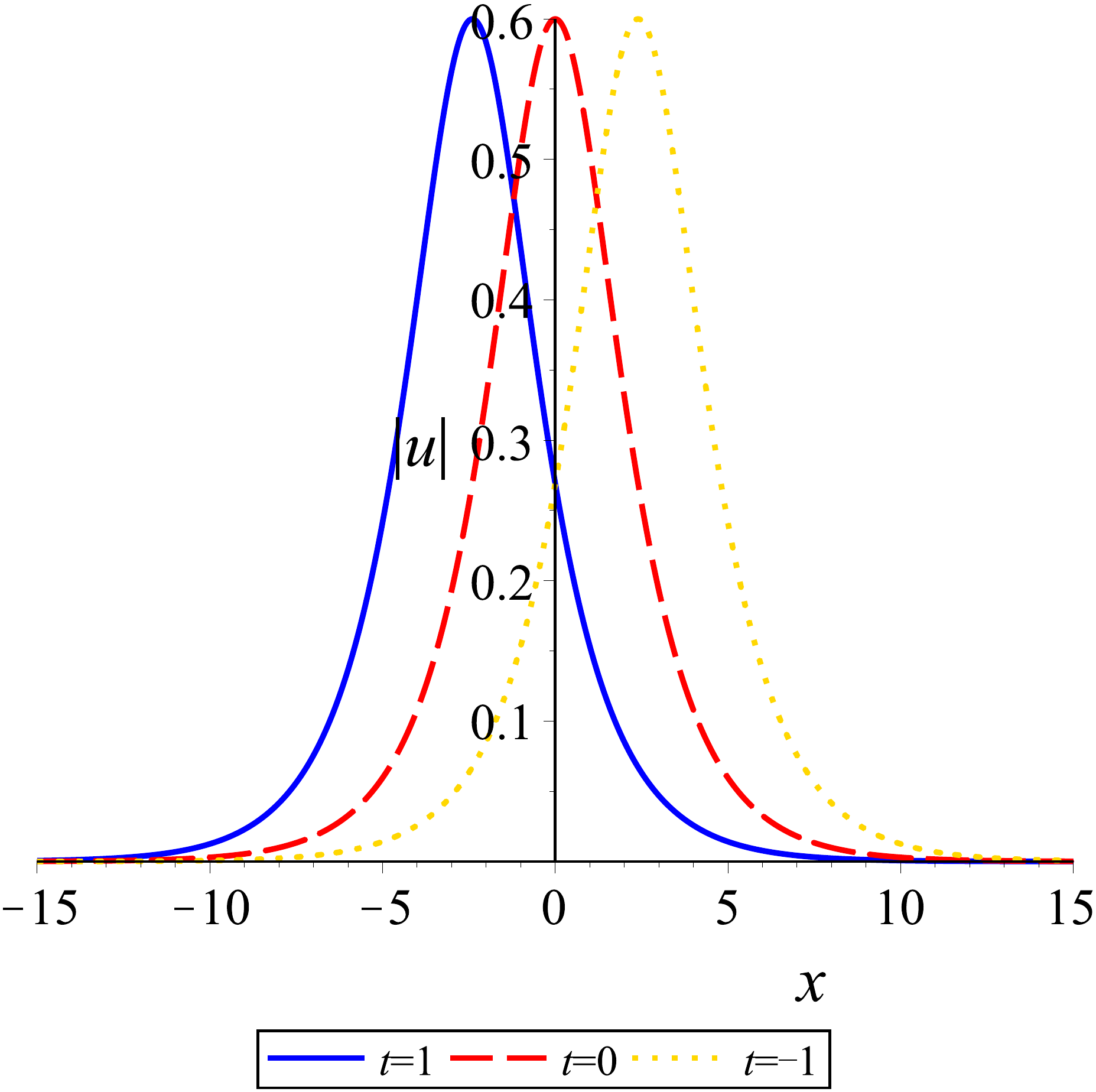}}}
\caption{Profiles of one-soliton solution (29) with $a_{{1}}=1,b_{{1}}= 0.5,\varsigma _{{11}}= 0.2,\varsigma _{{12}}=
 0.3,\xi_{{1}}=0,k=1,\alpha_{{1}}=1,\alpha_{{2}}=1,\alpha_{{3}}=1,y=0$. (a) three-dimentional plot; (b) $x$-curves.}
\end{center}
\end{figure}

(ii) When $N=2$, two-soliton solution is gained as
\begin{equation}
u(x,y,t)=-\frac{2\big(b _{1}{{a}_{1}^{*}}{{m}_{22}}{{{e}}^{\vartheta_{1}^{*}-{{\vartheta}_{1}}}}-b_{1}{a_{2}^{*}}{{m}_{12}}{{{e}}^{\vartheta_{2}^{*}-{{\vartheta }_{1}}}}-b _{2}{a_{1}^{*}}{{m}_{21}}{{{e}}^{\vartheta_{1}^{*}-{{\vartheta}_{2}}}}+b_{2}{{a }_{2}^{*}}{{m}_{11}}{{{e}}^{\vartheta_{2}^{*}-{{\vartheta}_{2}}}}\big)}{{{m}_{11}}{{m}_{22}}-{{m}_{12}}{{m}_{21}}},
\end{equation}
where
$$\begin{aligned}
&{{m}_{11}}=\frac{1}{{{\varsigma}_{1}}-\varsigma_{1}^{*}}{\big({{\left| {a_{1}} \right|}^{2}}{{{e}}^{\vartheta_{1}^{*}+{{\vartheta}_{1}}}}+{{\left| {b_{1}} \right|}^{2}}{{{e}}^{-\vartheta _{1}^{*}-{{\vartheta}_{1}}}}\big)},\quad
{{m}_{12}}=\frac{1}{{{\varsigma}_{2}}-\varsigma _{1}^{*}}{\big(a_{1}^{*}{a_{2}}{{{e}}^{\vartheta_{1}^{*}+{{\vartheta}_{2}}}}+b_{1}^{*}{b_{2}}{{{e}}^{-\vartheta_{1}^{*}-{{\vartheta }_{2}}}}\big)}, \\
&{{m}_{21}}=\frac{1}{{{\varsigma}_{1}}-\varsigma_{2}^{*}}{\big(a_{2}^{*}{a_{1}}{{{e}}^{\vartheta_{2}^{*}+{{\vartheta}_{1}}}}+b_{2}^{*}{b_{1}}{{{e}}^{-\vartheta_{2}^{*}-{{\vartheta }_{1}}}}\big)},\quad
{{m}_{22}}=\frac{1}{{{\varsigma}_{2}}-\varsigma_{2}^{*}}{\big({{\left| {a_{2}} \right|}^{2}}{{{e}}^{\vartheta_{2}^{*}+{{\vartheta}_{2}}}}+{{\left| {b_{2}} \right|}^{2}}{{{e}}^{-\vartheta _{2}^{*}-{{\vartheta}_{2}}}}\big)},
\end{aligned}$$
and
${{\vartheta}_{\iota}}=i{{\varsigma}_{\iota}}(x+ky)-i\alpha_{4}\varsigma_{\iota}^{2}t,{\varsigma}_{\iota}={\varsigma_{\iota 1}}+i{\varsigma_{\iota 2}},\iota=1,2$.
Under assumptions of $a_{1}=a_{2},{b_{1}}={b_{2}}=1$ and ${{\left| {a_{1}}\right|}^{2}}={{{e}}^{2{{\xi }_{1}}}}$, then the solution (30) is of the form
\begin{equation}
u(x,y,t)=-\frac{2\big({{a}_{1}^{*}}{{m}_{22}}{{{e}}^{\vartheta_{1}^{*}-{{\vartheta}_{1}}}}-{a_{2}^{*}}{{m}_{12}}{{{e}}^{\vartheta_{2}^{*}-{{\vartheta }_{1}}}}-{a_{1}^{*}}{{m}_{21}}{{{e}}^{\vartheta_{1}^{*}-{{\vartheta}_{2}}}}+{{a }_{2}^{*}}{{m}_{11}}{{{e}}^{\vartheta_{2}^{*}-{{\vartheta}_{2}}}}\big)}{{{m}_{11}}{{m}_{22}}-{{m}_{12}}{{m}_{21}}},
\end{equation}
where
$$
\begin{aligned}
&{{m}_{11}}=-\frac{i}{{{\varsigma}_{12}}}{{{e}}^{{{\xi }_{1}}}}\cosh(\vartheta_{1}^{*}+{{\vartheta}_{1}}+{{\xi }_{1}}),\quad {{m}_{12}}=\frac{2{{{e}}^{{{\xi }_{1}}}}}{({{\varsigma}_{21}}-{{\varsigma}_{11}})+i({{\varsigma}_{12}}+{{\varsigma}_{22}})}\cosh(\vartheta_{1}^{*}+{{\vartheta}_{2}}+{{\xi }_{1}}),\\
&{{m}_{22}}=-\frac{i}{{{\varsigma}_{22}}}{{{e}}^{{{\xi }_{1}}}}\cosh(\vartheta_{2}^{*}+{{\vartheta}_{2}}+{{\xi }_{1}}),\quad {{m}_{21}}=\frac{2{{{e}}^{{{\xi }_{1}}}}}{({{\varsigma}_{11}}-{{\varsigma}_{21}})+i({{\varsigma}_{12}}+{{\varsigma}_{22}})}\cosh(\vartheta_{2}^{*}+{{\vartheta}_{1}}+{{\xi }_{1}}).
\end{aligned}
$$

Below, we would like to examine two types of behaviors between two solitons:

Case 1. Two solitons propagate at the different speeds.
The parameters in (31) are selected as
$a_{1}=1,a_{2}=1,\varsigma _{11}=0.1,\varsigma _
{12}=0.3,\varsigma _{21}=0.3,\varsigma _{22}=0.5,\xi_{1}=0,k=1,\alpha_{1}=1,\alpha_{2}=1,\alpha_{3}=1,y=0,$
which can guarantee the different velocities of two solitons.
The corresponding solution can be computed directly and displayed by figure. Fig. 2(a) exhibits the localized structure of this solution in three-dimensions. The collision between two solitons occurs as shown in Fig. 2(b),
where two solitons propagate together towards the negative direction of the $x$-axis. The soliton with a larger amplitude moves much faster than the soliton with a smaller amplitude, and the larger soliton gradually overtakes the smaller as time goes on. When $t=0$, the amplitude superposition for two solitons reaches maximum value. And their interaction is elastic. The spatial structure of two solitons will be changed accordingly if we take other values for the parameters.
\begin{figure}
\begin{center}
\subfigure[]{\resizebox{0.36\hsize}{!}{\includegraphics*{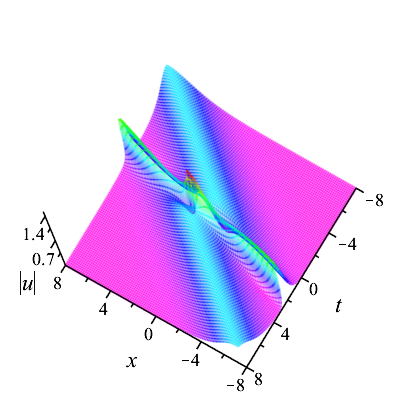}}}
\subfigure[]{\resizebox{0.32\hsize}{!}{\includegraphics*{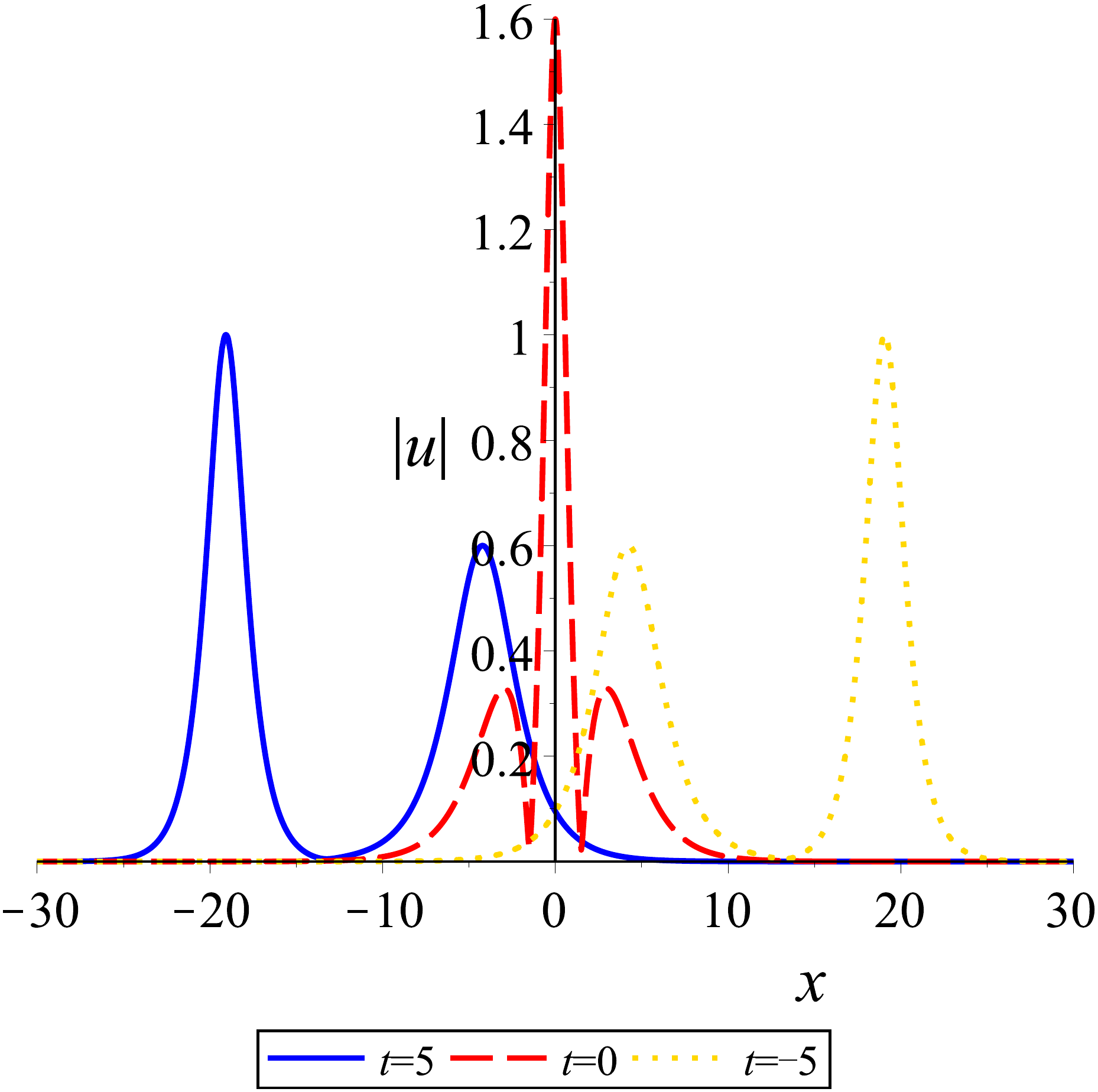}}}
\caption{Profiles of two-soliton solution (31) with $a_{1}=1,a_{2}=1,\varsigma _{11}=0.1,\varsigma _
{12}=0.3,\varsigma _{21}=0.3,\varsigma _{{22}}=0.5,\xi_{1}=0,k=1,\alpha_{1}=1,\alpha_{2}=1,\alpha_{3}=1,y=0$. (a) three-dimensional plot; (b) $x$-curves.}
\end{center}
\end{figure}

Case 2. Two solitons move at the equal speeds.
Specifying the solution parameters in (31) as
$a_{1}=1,a_{2}=1,\varsigma _{12}=0.3,\varsigma _{22}=0.5,\xi_{1}=0,k=1,\alpha_{1}=1,\alpha_{2}=1,\alpha_{3}=1,y=0$ as well as $\varsigma _{11}=\varsigma _{21}=0$ or $\varsigma _{11}=\varsigma _{21}=0.1$,
the corresponding solution can be worked out and demonstrated in Fig. 3.
Concerning this case, two solitons are spatial localization and stay together in traveling, namely, they are in a bound state.
According to these parameter values, the wave speed for two solitons moving from right to left along the $x$-axis is -1.2 in Fig. 3(c). It can be observed that when the solitons propagate, the amplitude function is periodic in oscillating as time goes on. And this solution represents a breather.
\begin{figure}
\begin{center}
\subfigure[]{\resizebox{0.35\hsize}{!}{\includegraphics*{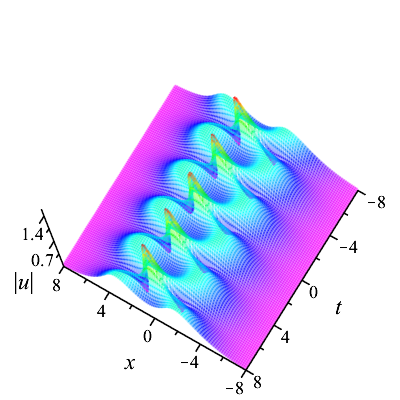}}}
\subfigure[]{\resizebox{0.31\hsize}{!}{\includegraphics*{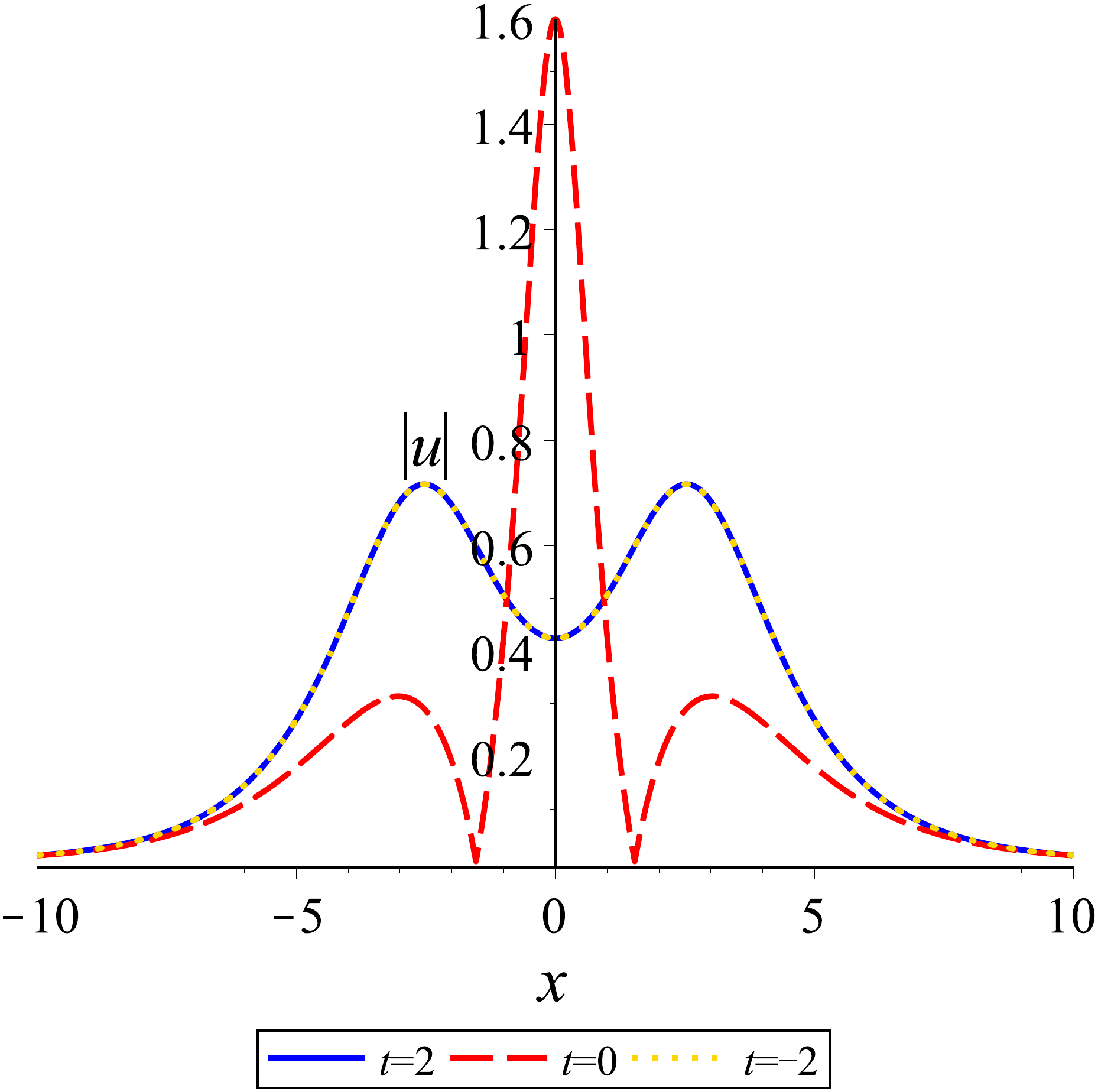}}}
\subfigure[]{\resizebox{0.35\hsize}{!}{\includegraphics*{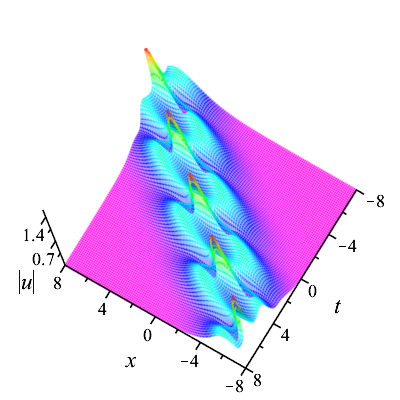}}}
\subfigure[]{\resizebox{0.31\hsize}{!}{\includegraphics*{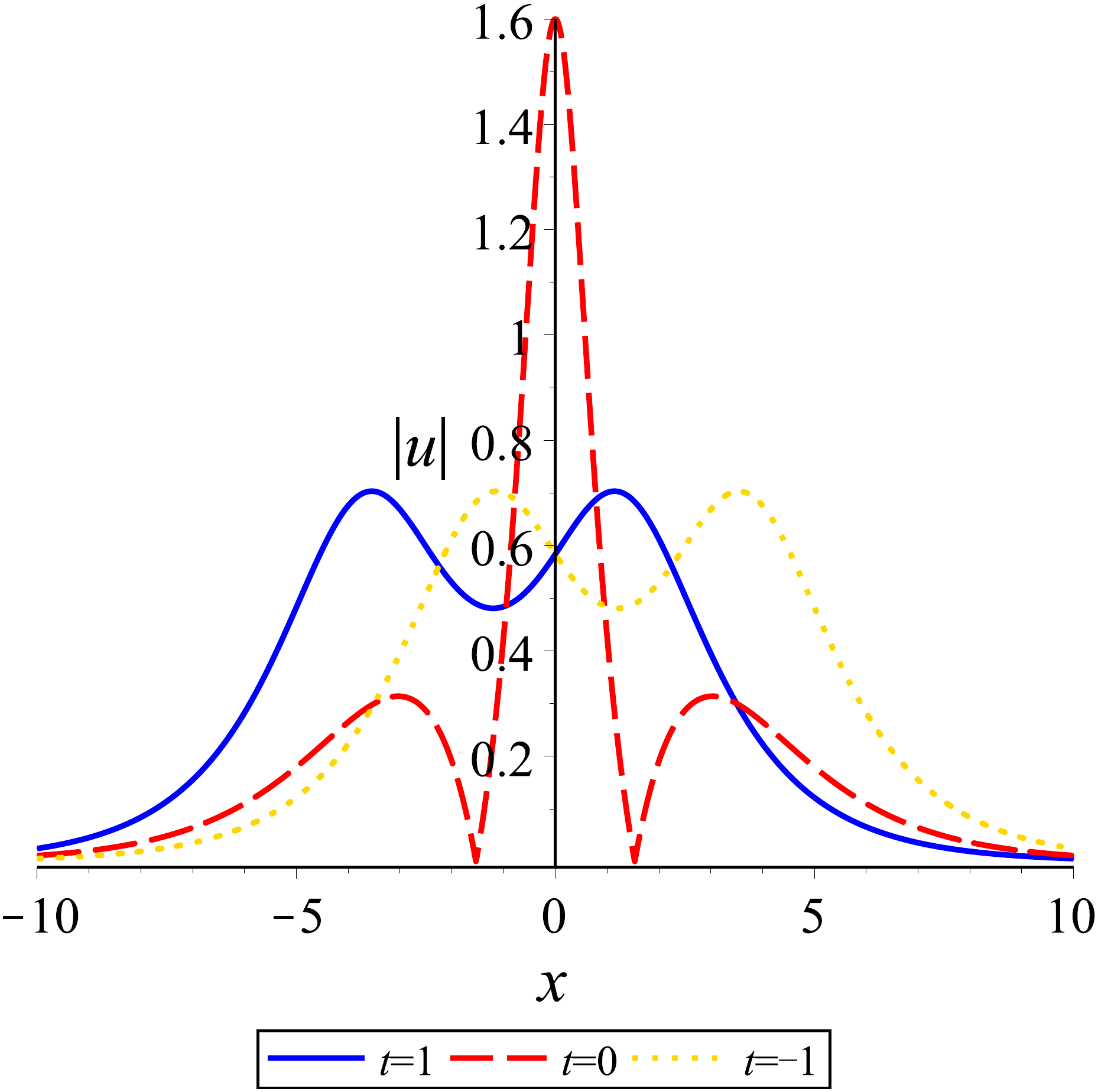}}}
\caption{Profiles of two-soliton solution (31) with $a_{1}=1,a_{2}=1,\varsigma _{12}=0.3,\varsigma _{22}=0.5,\xi_{1}=0,k=1,\alpha_{1}=1,\alpha_{2}=1,\alpha_{3}=1,y=0
$. (a) three-dimensional plot with $\varsigma _{11}=\varsigma _{21}=0$; (b) $x$-curves in Fig. 3(a); (c) three-dimensional plot with $\varsigma _{11}=\varsigma _{21}=0.1$; (d) $x$-curves in Fig. 3(c).}
\end{center}
\end{figure}
\section{Conclusion}
In summary, this paper mainly presents an application of the Riemann-Hilbert technique to the Heisenberg ferromagnetic spin chain equation in (2+1)-dimensions which models nonlinear wave propagation in ferromagnetic spin chain. As a result, the general multi-soliton solutions to the studied equation were attained. Through choosing suitable values for the relevant parameters, the localization in three-dimensions and dynamics in two-dimensions of one- and two-soliton solutions were depicted with the Maple plot tool.

\newpage

\end{CJK*}  

\addcontentsline{toc}{chapter}{References}
\begin{thebibliography}{99}\footnotesize
\itemsep=-3pt plus.2pt minus.2pt   

\bibitem{1} F.J. Yu, L. Li, Inverse scattering transformation and soliton stability for a nonlinear Gross--Pitaevskii equation with external potentials,
Appl. Math. Lett. 91 (2019) 41--47.
\bibitem{2} S. Zhang, C. Tian, W.Y. Qian, Bilinearization and new multisoliton solutions for the (4+1)-dimensional Fokas equation,
Pramana 86 (6) (2016) 1259--1267.
\bibitem{3} Y.S. Tao, J.S. He, Multisolitons, breathers, and rogue waves for the Hirota equation generated by the Darboux transformation, Phys. Rev. E 85 (2012) 026601.
\bibitem{4} X. L\"u, W.X. Ma, C.M. Khalique, A direct bilinear B\"acklund transformation of a (2+1)-dimensional Korteweg-de Vries-like model, Appl. Math. Lett. 50 (2015) 37--42.
\bibitem{5} D.S. Wang, S.J. Yin, Y. Tian, Y.F. Liu, Integrability and bright soliton solutions to the coupled nonlinear Schr\"odinger equation with higher-order effects, Appl. Math. Comput. 229 (2014) 296--309.
\bibitem{6} D.S. Wang, X.L. Wang, Long-time asymptotics and the bright $N$-soliton solutions of the Kundu--Eckhaus equation via the Riemann-Hilbert approach, Nonlinear Anal. Real World Appl. 41 (2018) 334--361.
\bibitem{7} W.X. Ma, Riemann-Hilbert problems and N-soliton solutions for a coupled mKdV system, J. Geom. Phys. 132 (2018) 45--54.
\bibitem{8} Z.Z. Kang, T.C. Xia, Construction of multi-soliton solutions of the $N$-coupled Hirota equations in an optical fiber, Chin. Phys. Lett. 36 (11) (2019) 110201.
\bibitem{9} M.M. Latha, C.C. Vasanthi, An integrable model of (2+1)-dimensional Heisenberg ferromagnetic spin chain and soliton excitations, Phys. Scr. 89
(2014) 065204.
\bibitem{11} H. Triki, A.M. Wazwaz, New solitons and periodic wave solutions for the (2+1)-dimensional Heisenberg ferromagnetic spin chain equation, J. Electromagn. Waves Appl. 30 (2016) 788--794.
\bibitem{14} M. Inc, A.I. Aliyu, A. Yusuf, D. Baleanu, Optical solitons and modulation instability analysis of an integrable model of (2+1)-dimensional Heisenberg ferromagnetic spin chain equation, Superlattice. Microstruct. 112 (2017) 628--638.
\bibitem{14} G. Tang, S. Wang, G. Wang, Solitons and complexitons solutions of an integrable model of (2+1)-dimensional Heisenberg ferromagnetic spin chain, Nonlinear Dyn. 88 (2017) 2319--2327.
\bibitem{11} Y.L. Ma, B.Q. Li, Y.Y. Fu, A series of the solutions for the Heisenberg ferromagnetic spin chain equation, Math. Methods Appl. Sci. 41 (2018) 3316--3322.
\bibitem{11} T.A. Sulaiman, T. Akt\"urk, H. Bulut, H.M. Baskonus, Investigation of various soliton solutions to the Heisenberg ferromagnetic spin chain equation, J. Electromagn. Waves Appl. 32 (9) (2018) 1093--1105.
\bibitem{11} H. Bulut, T.A. Sulaiman, H.M. Baskonus, Dark, bright and other soliton solutions to the Heisenberg ferromagnetic spin chain equation, Superlattice. Microstruct. 123 (2018) 12--19.
\bibitem{10} B.Q. Li, Y.L. Ma, Lax pair, Darboux transformation and Nth-order rogue wave solutions for a (2+1)-dimensional Heisenberg ferromagnetic spin chain equation, Comput. Math. Appl. 77 (2019) 514--524.
\bibitem{14} X.X. Du, B. Tian, Y.Q. Yuan, Z. Du, Symmetry reductions, group-invariant solutions, and conservation laws of a (2+1)-dimensional nonlinear Schr\"odinger equation in a Heisenberg ferromagnetic spin chain, Ann. Phys. 531 (2019) 1900198.
\bibitem{14} A.R. Seadawy, N. Nasreen, D. Lu, M. Arshad, Arising wave propagation in nonlinear media for the (2+1)-dimensional Heisenberg ferromagnetic spin chain dynamical model, Physica A 538 (2020) 122846.
\bibitem{14} M.S. Osman, K.U. Tariq, A. Bekir, A. Elmoasry, N.S. Elazab, M. Younis, M. Abdel-Aty, Investigation of soliton solutions with different wave structures to the (2+1)-dimensional Heisenberg ferromagnetic spin chain equation, Commun. Theor. Phys. 72 (2020) 035002.
\bibitem{12} B.Q. Li, Interaction behaviors between breather and rogue wave in a Heisenberg ferromagnetic equation, Optik 227 (2021) 166101.
\bibitem{13} Y.L. Ma, Lump wave phase transition for the (2+1)-dimensional Heisenberg ferromagnetic spin chain equation, Optik 231 (2021) 166505.
\bibitem{14} D. Yang, Traveling waves and bifurcations for the (2+1)-dimensional Heisenberg ferromagnetic spin chain equation, Optik 248 (2021) 168058.
\bibitem{11} S. Sahoo, A. Tripathy, New exact solitary solutions of the (2+1)-dimensional Heisenberg ferromagnetic spin chain equation, Eur. Phys. J. Plus 137 (2022) 390.
\bibitem{14} K.S. Nisar, M. Inc, A. Jhangeer, M. Muddassar, B. Infal, New soliton solutions of Heisenberg ferromagnetic spin chain model, Pramana 96 (2022) 28.
\bibitem{15} J.K. Yang, Nonlinear Waves in Integrable and Nonintegrable Systems. SIAM, Philadelphia (2010).
\end{thebibliography}
\end{document}